\documentclass[conference]{IEEEtran}
\usepackage{times}

\usepackage[numbers]{natbib}
\usepackage{multicol}
\usepackage[bookmarks=true]{hyperref}

\usepackage{multirow,array}
\usepackage{subfig}
\usepackage{amsmath} 
\usepackage{amssymb}  
\usepackage{svg}

\usepackage{comment} 

\DeclareMathOperator*{\argmin}{arg\,min}

\usepackage{booktabs}

\usepackage{amsthm}
\newtheorem*{definition}{Definition}

\title{Resolving Conflict in Decision-Making for Autonomous Driving}

\author{\authorblockN{Jack Geary}
\authorblockA{School of Informatics\\
University of Edinburgh, UK\\
jack.geary@ed.ac.uk}
\and
\authorblockN{Subramanian Ramamoorthy}
\authorblockA{School of Informatics\\
University of Edinburgh, UK\\
Five AI Ltd., UK\\
s.ramamoorthy@ed.ac.uk}
\and
\authorblockN{Henry Gouk}
\authorblockA{School of Informatics\\
University of Edinburgh, UK\\
henry.gouk@ed.ac.uk}}

\date{March 2021}

\begin{document}

\maketitle

\begin{abstract}

Recent work on decision making and planning for autonomous driving has made use of game theoretic methods to model interaction between agents. We demonstrate that methods based on the Stackelberg game formulation of this problem are susceptible to an issue that we refer to as conflict. Our results show that when conflict occurs, it causes sub-optimal and potentially dangerous behaviour. In response, we develop a theoretical framework for analysing the extent to which such methods are impacted by conflict, and apply this framework to several existing approaches modelling interaction between agents. Moreover, we propose Augmented Altruism, a novel approach to modelling interaction between players in a Stackelberg game, and show that it is less prone to conflict than previous techniques. Finally, we investigate the behavioural assumptions that underpin our approach by performing experiments with human participants. The results show that our model explains human decision-making better than existing game-theoretic models of interactive driving.
\end{abstract}

\section{Introduction}
\begin{figure}[!ht]
    \centering
    \subfloat[\label{intro_figure_image}]{%
        \includegraphics[height=0.25\textheight]{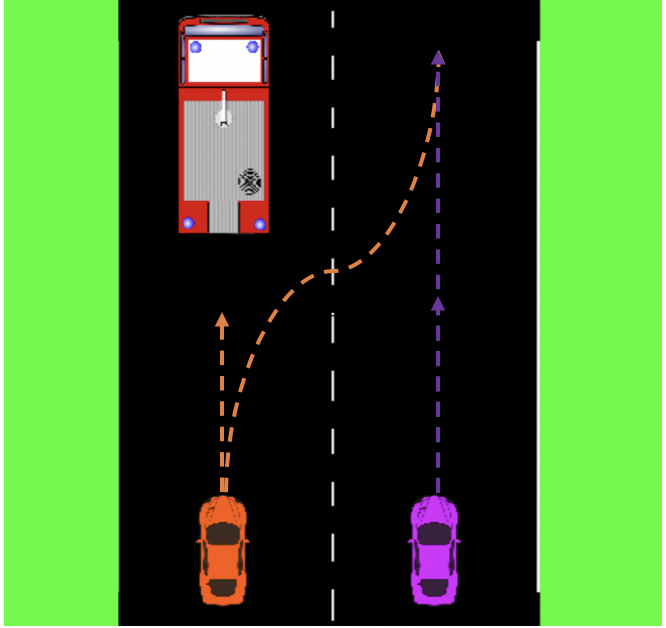}}
    \newline
    \begin{flushleft}
        \subfloat[\label{intro_figure_table}]{%
            \begin{tabular}{cc|c|c|}
                & \multicolumn{1}{c}{} & \multicolumn{2}{c}{$C2$}\\
                & \multicolumn{1}{c}{} & \multicolumn{1}{c}{$Y$}  & \multicolumn{1}{c}{$C$} \\\cline{3-4}
                \multirow{2}*{$C1$}  & $LCB$ & $(-\infty,-\infty)$ & $(0,1)$ \\\cline{3-4}
                & $LCA$ & $(1,0)$ & $(-\infty,-\infty)$ \\\cline{3-4}
            \end{tabular}}
    \end{flushleft}
\caption{(a): Motivating Example; Car 1 (Orange) is in the left lane approaching a stopped fire truck. Car 2 (Purple) is in the adjacent lane. Dotted lines depict the options available to each vehicle; Car 1 must change lanes ahead of Car 2 (LCA), or behind Car 2 (LCB). Car 2 can either maintain their current velocity (C) or yield (Y) to allow Car 1 to merge onto the lane. (b) Reward matrix associated with the motivating example; Car 2 would prefer to continue, and Car 1 would prefer to change lanes ahead of Car 2. Diagonal entries in table represent states where the cars collide or neither satisfies their objective, which is mutually undesirable.}
\label{intro_figure}
\end{figure}

A wealth of previous research on autonomous driving has employed overly simplistic models of other agents' behaviour, such as assuming constant velocity~\citep{Eiras2020,sadigh2015,Bender2015} or that vehicles can only execute a small number of fixed trajectories for a given type of manoeuvre~\citep{hermes2009,Ward2017}. These assumptions can lead to unexpected and undesirable behaviour, like failing to change lanes or merge onto a highway, as the models do not anticipate a driver's inclination to make space to complete a manoeuvre if it is initiated. The emerging field of interactive planning and decision-making for autonomous driving aims to address these problems by building  models that account for the autonomous vehicle's (AV) ability to induce behaviours in other vehicles.~\citep{sadigh2016,Sadigh2018}. For example, the scenario presented in Figure~\ref{intro_figure_image}; the AV (orange) is approaching a parked fire truck with a vehicle (purple) occupying the adjacent lane. Operating under the assumption  that the purple car will continue to move at a constant velocity, the AV has no choice but to slow down and wait until the right lane is empty in order to continue. In contrast, an AV that is capable of anticipating how an interactive driver would respond to its actions might decide on a different course of action. For example, \citet{Sadigh2018} demonstrate an interactive planner that enables the AV to choose behaviours in order to determine whether the purple car will slow down to allow it to change lanes in front.

A common approach to trajectory planning is to develop hierarchical systems, where a discrete decision-maker decides on intents or approximate trajectories for the AV, and a lower level motion planning system chooses continuous actions that realise these decisions ~\citep{fisac2019,albrecht2020,Eiras2020,Cunningham2015}. Incorporating models of interaction based on Stackelberg games into AV planning and decision-making systems has become a popular way of improving upon na\"ive models that neglect the interactive behaviour of other drivers. However, while solving Stackelberg games is computationally easier than alternative game theoretic formulations, the solution method presumes that players (i.e., the AV and other driver) take the role of either leader of follower, with the leader selecting their action first and the follower selecting their best response. In practice, without a means of direct communication and agreed upon protocol, it is impractical for two vehicles to dynamically allocate the roles of leader and follower. Given that the roles of leader and follower are ambiguous, the assumptions required for a Stackelberg game do not always hold in autonomous driving scenarios. This begs the question: will the violated assumptions result in undesirable behaviour?

We use the term \emph{conflict} to refer to a class of situations where ambiguity over who takes the roles of leader and follower leads to unexpected behaviour. We demonstrate that such conflicts can result in sub-optimal or unsafe behaviour, such as both vehicles coming to a stop in an attempt to yield to each other, or both vehicles aggressively trying to move into the same space. In addition, we propose a metric---the Area of Conflict (AoC)---for the rate at which conflict is expected to occur in a given Stackelberg game-based model. This metric can be seen as a measure of robustness to the level of aggressiveness or passiveness exhibited by other agents in the driving environment. Using the insight provided by our conflict analysis we define Augmented Altruism, a model for interactive decision-making that is less susceptible to conflict than previous approaches. Our approach is based on the concept of altruism~\citep{andreoni1993} from game theory, which we extend to account for reciprocal altruistic considerations of other agents. Using data collected from both simulated and human experiments, we demonstrate the impact conflict can have on driving outcomes in interactive settings. Our results also provide evidence indicating that Augmented Altruism more accurately models human behaviour than current methods. 

The key contributions in this work are as follows:
\begin{itemize}
    \item Identification of a shortcoming in previous work on interactive planning and decision-making for autonomous driving that we refer to as conflict.
    \item The introduction of a metric, Area of Conflict (AoC), for measuring the incidence of conflict for a pair of decision-makers in a Stackelberg game formulation.
    \item We propose Augmented Altruism, a novel method for decision-making that is provably less prone to conflict than other methods in literature for reasonable reward values.
\end{itemize}
\section{Related Work}
Our approach is an extension of the concept of Altruism, as found in Game Theory literature. \citet{andreoni1993} presents the idea of altruism being a scalar value, $\alpha$, that multiplies or adds with the rewards of the interacting agents to influence an agent's decision-making by the potential payoffs to the other agents. In the work, \citeauthor{andreoni1993} provide three distinct models for altruism: pure, duty and reciprocal. The definition used in this work most aligns with the definition for pure altruism. \cite{Bansal2018} introduce selfishness, an equivalent, but opposite concept to our proposed definition, which is used for collaborative automotive planning. \cite{during2014} presents a method for decision-making for AVs that corresponds to the pure altruism model, with an altruism coefficient of $1$. 

Similar to altruism in Game Theory, there is Social Value Orientation (SVO) in the fields of psychology and behavioural economics \citep{mcclintock1989}. SVO can be used as an indicator for a person's reward allocation preferences in coordination tasks. Unlike altruism, which is restricted to depicting egoistical or prosocial behaviours, SVO can also identify malicious or masochistic behaviours in decision-makers. \citet{Schwarting2019} implement a version of SVO where the planning agent's reward, and the rewards of the other agents, are weighted according to the planning agent's SVO value. We use a similar approach to incorporate altruism in this work. As in \cite{sadigh2016}, the authors model the interaction as a Stackelberg game and demonstrate that SVO can be used to augment lane merge prediction. In contrast, in this work we use our proposed method as a decision-making model. In \cite{Schwarting2019} the SVO model requires access to a single, accurate reward function to model the behaviour of any other agent, which is learnt offline. In our work we weaken this requirement, using Game Theoretic methods to choose the most appropriate reward function to model the behaviour of other vehicles.

\citet{sadigh2016} present an approach to interactive planning, in which a reward function is learnt from human data, and then the structure of the reward function is exploited in order to identify optimal interactive actions \citep{sadigh2016, Sadigh2018}. \citet{fisac2019} proposes a similar approach, although they do not rely on knowing the structure of the reward function in their planning, only that the reward function is known and accurate. These works formulate the interaction problem as a two-person Stackelberg game \citep{von1934} with a dense reward, assuming that the other agents in the environment abide by the leader-follower hierarchy specified.

All of these approaches to modelling interaction with Stackelberg games require the roles of leader and follower to be agreed upon beforehand in order to avoid undesirable behaviour. Our theoretical analysis shows that the method we propose is less vulnerable to this ambiguity than these previous approaches.
\section{Conflict in Stackelberg Games}
In a two-person Stackelberg game one player takes the role of the leader and the other the role of the follower. The leader chooses the action that maximises their reward, according to some known reward matrix, under the assumption that the follower will behave optimally with respect to the leader's choice. For example, using the reward matrix given in Figure \ref{intro_figure_table}, if $C1$ were the leader then they would choose to lane change ahead of $C2$ (and get reward of 1) anticipating that the follower, $C2$ will respond by yielding (and get a reward of 0). However, if $C2$ were the leader instead, they would choose to continue (and get reward 1) and $C1$ would be forced to decelerate. Thus, if both cars independently assume the role of leader, $C1$ will attempt a lane change and $C2$ will not yield. This example shows that,if it has not been agreed in advance which agent is the leader and which agent is the follower, the uncertainty can result in the agents computing conflicting equilibria for the game. In the case of autonomous driving, without any means of direct communication, no such agreement can be reached. We define conflict as follows:

\begin{definition}
    Conflict occurs in a Stackelberg game when a lack of agreement about who takes the role of leader and who takes the role of follower leads to the players arriving at different pure strategy equilibria.
\end{definition}

Conflicts can be resolved in one of two ways: (i) enabling players to negotiate the roles of leader and follower before playing the game, or (ii) designing the reward such that the players will arrive at the same pure strategy equilibrium irrespective of who takes the role of leader or follower.

Conflicts can be problematic as they can result in unforeseen catastrophic situations, as shown in the previous example. Therefore it is important that the decision-making method used by an AV has as low an incidence of conflict possible, so that it is robust to the aggressiveness of other agents. In our example the reward matrix indicates that the players are in conflict, and there is no clear way to resolve it.
\section{Impact of Conflict}
In this section we show how conflict can impact the ability of an agent to accomplish its objective in a safe and timely fashion. We consider the four possibilities; both agents plan assuming they are the leader, both agents agree Car 1 is the leader, both agents agree Car 2 is the leader, and both agents think the other is the leader. When the agents do not agree, which can occur if there is conflict in the game, they will compute different equilibria to the game. We use a simulated highway lane change driving scenario to demonstrate that this conflict can cause sub-optimal behaviour, resulting in longer objective completion times.

\subsection{Experimental Setup}
Each agent $i$ has a finite set of intentions, $A_{i}$, with each intention associated with a cost function, $J_{a}: X^{T} \rightarrow \mathbb{R}, a \in A_{i}$, where $X$ is the state space and $T$ is the trajectory duration. An equilibrium to a game, $(a_{i},a_{-i})$, defines a cost function, $J_{i} = J_{a_{i}} + J_{a_{-i}}$. Each agent $i$ uses the cost function $J_{i}$ compute an optimal trajectory to achieve the joint intentions. We treat the problem as a receding horizon optimal control problem, and use a Model Predictive Control (MPC) planning scheme to solve it.

The vehicle's state is defined by its $x$ and $y$ position as well as its velocity and heading; $\vec{x} = [x,y,v,\theta] \in X$. Vehicles can control their linear acceleration, $a$, and angular acceleration, $\omega$. Control inputs are of the form $\vec{u}=(a,\omega)$. The standard bicycle model, $F: X \times \mathbb{R}^{2} \rightarrow X$ \citep{kong2015}, is used to define the vehicle dynamics. Vehicles are rectangular with length $L=4.6$ metres and width $W=2$ metres. Each lane is $4$ metres wide.

At every iteration of the MPC, each agent generates a trajectory $x^{*}_{i} = \{\vec{x}_{i}^{t}\}_{t=0}^{T}, u^{*}_{i} = \{\vec{u}_{i}^{t}\}_{t=0}^{T}$  such that 
\begin{equation*}
\begin{aligned}
x^{*}_{i},u^{*}_{i},x^{*}_{-i}, u^{*}_{-i}& = \argmin_{x_{i},u_{i},x_{-i},u_{-i}} \quad  J_{i}(x_{i},u_{i},x_{-i},u_{-i})\\
\textrm{s.t. } & \vec{x}_{j}^{t+1} = F(\vec{x}_{j}^{t},\vec{u}_{j}^{t})\quad \forall 0 \leq t < T , j \in \{i,-i\} \\
              &  g(x_{j}) \leq 0, j \in \{i,-i\} \\
              &  h(u_{j}) \leq 0, j \in \{i,-i\}  \\
              & \frac{(\vec{x}_{i}^{t}[0]-\vec{x}_{-i}^{t}[0])^{2}}{a^2} + \frac{(\vec{x}_{i}^{t}[1]-\vec{x}_{-i}^{t}[1])^{2}}{b^2} \geq 1.
\end{aligned}
\end{equation*}
where the index $-i$ identfies the agent not indexed by $i$. For each agent $i$ this method also produces $x^{*}_{-i}, u^{*}_{-i}$  which is the trajectory agent $-i$ will follow, presuming their trajectory is also optimal with respect to $J_{i}$ (this will be the case if the agents are not in conflict as they will have computed the same equilibrium). These values are discarded by agent $i$. The vehicles then follow their respective optimal trajectories for two timesteps before replanning. We presume that at the time of planning the agents have perfect awareness of each other's states. A lookahead horizon of $T=4$ seconds is used, with a timestep, $\text{dt}=.2$ seconds. The experiment finishes when both agents have completed their objective, or if the trajectory duration exceeds $10$ seconds.

Constraints, $g$, are applied to the physical state of each agent; these ensure the vehicles stay on the road, and that each vehicle's velocity does not exceed $15m/s$. Action constraints, $h$, enforce that acceleration is in the range of $[-9m/s^{2},3m/s^{2}]$, and the angular acceleration is in the range $[-1deg/s^{2},1deg/s^{2}]$. The final constraint is an obstacle avoidance constraint, to ensure that the vehicles do not construct plans in which collisions occur. An ellipse with semi-major and semi-minor axes $a$ and $b$ is fit around agent $i$'s position, preventing the vehicles from getting too close. In all experiments $a=W + \delta$ and $b=L + \epsilon$ where $\epsilon$ and $\delta$ are small constants. An Interior Point Optimizer (IPOPT) method from the CasADi optimisation library \citep{Casadi} is used to solve the resulting non-linear programming problem.\

We record the time taken for both agents to complete their objectives, as measured by a boolean check based on the intention specified in the computed equilibrium. Trajectories that fail to complete both agent's objectives (e.g., due to a collision or frozen robot problem~\citep{Trautman2010}) are given a maximal score. The hypothesis is that conflict interferes with the ability to interact effectively, hence the time for the objectives to be completed will be longer for those sets of leader-follower assumptions that are in conflict. We evaluate this claim by staggering the starting positions of each vehicle and recording the completion times. The vehicles are displaced to a maximum displacement of $1.5$ vehicle lengths ($6.9$ metres). Empirically we observe that outside this range the cars cease to be interacting and can generally accomplish their objectives without influencing the other's behaviour.

\subsection{Simulated Lane Change Experiment}
Figure \ref{intro_figure_image} demonstrates the default setup for the lane change experiment; both cars start next to each other in adjacent lanes, travelling at the speed limit. Car 1 (Orange) can either choose to change lanes ahead (LCA) of Car 2 (Purple), or change lanes behind (LCB). Car 2 can either Yield (Y) to allow Car 1 to perform the lane change ahead, or else they can continue (C). Car 1 would prefer to merge ahead of Car 2, but Car 2 would prefer not to yield, which would prevent Car 1 from merging ahead (this is captured in the reward matrix defined in Figure \ref{intro_figure_table}). If Car 1 believes they are the leader, they will optimise for the equilibrium $(LCA,Y)$, whereas, using the same assumption, C2 will optimise $(LCB,C)$, resulting in the action pair $(LCA,C)$ being executed. As these equilibria do not match, we conclude the game is in conflict. 

All $J_{a}, a \in A_{i} \cup A_{-i}$ in this experiment have common features incentivising the vehicles to get into the right hand lane as early as possible, and to drive at the speed limit as much as possible. These features are weighted the same $\forall a \in A_{i} \cup A_{-i}$, so that neither agent's cost function prioritises their objective over the other's. The remaining feature penalises an agent $i$ for being ahead of agent $-i$ if agent $i$ is expected to yield. This is given as
\begin{equation*}
    f(\vec{x}_{i},\vec{x}_{-i}) = \max(\vec{x}_{i}[1]-\vec{x}_{-i}[1],0)
\end{equation*}
The weights of all features are fit experimentally, and all features have positive non-zero weight.\ 

\begin{figure}[t]
    \centering
    \subfloat[\label{fig:exp1_results}]{%
        \includegraphics[width=.9\columnwidth]{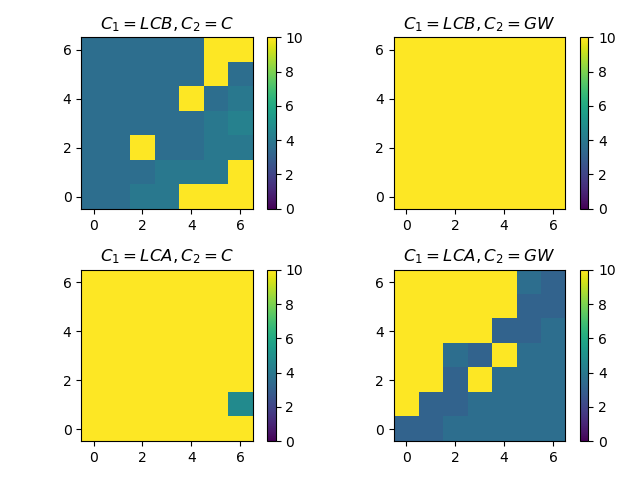}}
    \newline
        \subfloat[\label{table:exp1_results}]{%
        \begin{tabular}{llc}
             \toprule
            C1 Action   & C2 Action & Average Time (s)\\
            \midrule
            LCA         & C         & $9.89$ \\
            LCA         & Y        & $6.25$ \\
            LCB         & C         & $4.87$ \\
            LCB         & Y        & $10.0$   \\
            \bottomrule
        \end{tabular}}
    \caption{Results from Experiment 1. (a) Trajectory duration achieved in the staggered experiments; each plot depicts the outcomes for a different action combination. The x-axis of each plot gives offset of Car 1 (Orange), and the y-axis gives the offset for Car 2 (Purple) from the default position. (b) The average trajectory duration achieved over all the experiments for each action pair.}
\label{exp1_results}
\end{figure}

The results in Figure \ref{exp1_results} demonstrate that when both agents decide on the same equilibrium (e.g., in the case of $(LCA,Y)$ and $(LCB,C)$) the cars are consistently able to resolve the scenario cooperatively in less time than in conflict. If both cars presume the other car is the leader in the Stackelberg game, they both attempt to accommodate the other, effectively achieving the equilibrium $(LCB,Y)$, and they fail to complete their objectives. Similarly, when both cars presume they are the leader, they effectively execute the $(LCA,C)$ solution. In this case the cars almost always fail to complete the task safely. Even when the cars are not in conflict, if Car 2 started ahead of Car 1, and Car 1 is attempting to merge ahead, this can result in a collision, so conflict does not entirely eliminate the risk in the manoeuvre.

\section{Altruism and Area of Conflict}
\label{altruism_and_area_of_conflict}
Altruism-based techniques for decision making in game theory can be thought of as methods for transforming the reward matrix of a game. In this section we will define our variant of altruism~\citep{andreoni1993}, as well as an augmentation to the definition that accounts for the reciprocal considerations of other altruistic agents. We will also provide a definition for Area of Conflict, our proposed measure for the expected incidence of conflict in a Stackelberg game. 

\subsection{Altruism}
\begin{figure}[t]
    \centering
    \begin{tabular}{cc|c|c|}
        & \multicolumn{1}{c}{} & \multicolumn{2}{c}{$C$}\\
        & \multicolumn{1}{c}{} & \multicolumn{1}{c}{$B1$}  & \multicolumn{1}{c}{$B2$} \\\cline{3-4}
        \multirow{2}*{$R$}  & $A1$ & $(r_{111},r_{112})$ & $(r_{121},r_{122})$ \\\cline{3-4}
        & $A2$ & $(r_{211},r_{212})$ & $(r_{221},r_{222})$ \\\cline{3-4}
    \end{tabular}
\caption{General reward matrix}
\label{reward_matrix}
\end{figure}

We model the interactive driving decision-making problem as a static game played on a reward matrix, indexed by intentions, where each cell in the matrix contains the rewards received by each player if they each chose the corresponding intention combination.  Figure \ref{reward_matrix} presents a general reward matrix where if the row player, $R$, and column player $C$, chose intentions $A1$,$B1$ respectively, $R$ would receive a reward of $r_{111}$ and $C$ would receive a reward of $r_{112}$. The grid is $2 \times 2$ for demonstrative purposes and, in general, the grid can be of any size $M \times N$ where $M$ is the number of actions available to $R$ and $N$ is the number of actions available to $C$, and each cell contains a reward pair $(r_{mn1},r_{mn2})$. Unless the full index is required, we will refer to the row player's reward for a particular intention combination as $r_{1}$ and, correspondingly, $r_{2}$ for the column player's reward.

Pure Altruism, as defined in \citep{andreoni1993}, makes use of an altruism coefficient $\alpha$ to define the altruistic reward,
\begin{equation}
    r^{*}_{i} = r_{i} + \alpha r_{-i} \quad 0 \leq \alpha \leq 1,
\end{equation}
where $r^{*}_{i}$ is the effective reward agent $i$ uses to perform decision-making. If $\alpha=0$ then the agents are indifferent to one another and if the value is $1$ then the agents are cooperating in order to maximise the same reward, $r^{*}_{i} = r^{*}_{-i} =  r_{i} + r_{-i}$. 

As an alternative to Pure Altruism we propose an alternative definition for the altruistic reward,
\begin{equation}
    r^{*}_{i} = (1-\alpha_{i})r_{i} + \alpha_{i}r_{-i} \quad 0 \leq \alpha_{i} \leq 1.
    \label{eqn:altruism_definition}
\end{equation}
In this case each agent $i$ has their own individual altruism coefficient, $\alpha_{i}$. Scaling agent $i$'s reward in parallel with agent $-i$'s allows for more flexible behaviours as compared with Pure Altruism; if $\alpha_{i}=0$ then the agent is wholly egoistic, if $\alpha_{i}=1$ then the agent is wholly altruistic. To avoid confusion we will refer to the \citet{andreoni1993} altruism as ``pure altruism'' and our proposed definition as ``altruism''. We refer to models that depend on a single interaction parameter(e.g. Pure Altruism, SVO, Altruism) as ``egoistic'' interactive models. 

Extensive previous and ongoing work has been dedicated to estimating reward functions and interactive parameters \citep{albrecht2016,albrecht2019,albrecht2020,Schwarting2019,sadigh2016,landolfi2018,sun2018}. In this work we presume that the ``true'' reward matrix $\{(r_{mn1},r_{mn2})\}_{0 < m \leq M, 0 < n \leq N}$, and altruism values $\alpha_1,\alpha_2$ are known to both agents. Each agent can then, independently, construct the reward matrix $\{(r{*}_{mn1},r^{*}_{mn2})\}_{0 < m \leq M, 0 < n \leq N}$, which they will use to choose which intention to follow. 

\subsection{Augmented Altruism}
When attempting to identify equilibria in Game Theoretic problem formulations it is not uncommon to use iterative best response methods to compute the Nash Equilibrium \citep{vorobeychik2008}. In practise this involves each agent choosing an optimal action based on the optimal actions for the other agents in the previous iteration. This allows for the fact that an agent's choice of action can be affected by the choices made by other agents. If this process is repeated indefinitely, and it converges to a solution, then the solution achieved is a Nash Equilibrium \citep{bacsar1998}.

The altruism definition presented in Equation \ref{eqn:altruism_definition} neglects from consideration that $r_{-i}^{*}$ is the reward an altruistic agent $-i$ would receive from the interaction, not $r_{-i}$, and that awareness of this value could affect agent $i$'s preferences. But, by the same assumption, the value for $r_{-i}^{*}$ depends on the value of $r_{i}^{*}$. By treating the equations in Equation \ref{eqn:altruism_definition} as a system of equations, we can determine the steady state of the system, yielding what we refer to as ``augmented altruism'',
\begin{equation}
    r_{i}^{*} = \dfrac{(1-\alpha_{i})r_{i}+\alpha_{i}(1-\alpha_{-i})r_{-i}}{1-\alpha_{i}\alpha_{-i}} \quad i \in \{1,2\}
    \label{eqn:augmented_altruism_definition}
\end{equation}
This improves on our base altruism definition, as it is a computationally tractable method for accounting for both players altruism values when evaluating options, whereas egoistic models only account for the agent's own $\alpha$. We refer to models with this property as ``non-egoistic'' models. For a more detailed explanation and complete derivation, we direct the reader to the Appendix \ref{deriving_augmented_altruism} . 

\subsection{Area of Conflict}
Altruism can be used to resolve conflict scenarios; in the example in Figure \ref{intro_figure_table}, if $\alpha_{1}=1$, for instance, then $C1$ would get an effective reward of $0$ for attempting to cut ahead, and a reward of $1$ for decelerating and allowing $C2$ to proceed. However, altruism does not entirely eliminate conflict since $\alpha_{1}=1 \text{ and }\alpha_{2}=1$ also results in conflict. 

Let each agent choose an action according to $f_{I}:\mathbb{R}^{M \times N} \times [0,1] \times [0,1] \rightarrow \{A1,A2\} \times \{B1,B2\}$, a function parameterised by the altruism coefficients that maps from the reward matrix to the equilibrium of the corresponding Stackelberg game for some reward matrix $I$. The previous observation indicates that, for a given reward matrix, there are potentially regions in the parameter-space $[0,1] \times [0,1]$ that will always result in conflict. We call the total size of these regions the Area of Conflict (AoC). It is desirable to choose a decision-making method that minimises the AoC for a given reward matrix.

In the following derivations we will refer to the reward matrix defined in Figure \ref{reward_matrix}. Without loss of generality we will assume the cell $(A2,B1)$ is optimal for $R$ and $(A1,B2)$ is optimal for $C$. We further assume that there are no ambiguities in each agents' rewards, i.e.;
\begin{equation}
    \begin{split}
        r_{211}>&r_{121},r_{111},r_{221} \\
        r_{122}>&r_{212},r_{112},r_{222}.
    \end{split}
\end{equation}
It is clear that decision-making on the reward matrix with these constraints will result in conflict, regardless of the value of the parameters. Therefore it is vacuously true that if $I$ is the unchanged reward matrix, then the AoC of the resulting Stackelberg game is $1$~\citep{von1934}. We will use this value as a baseline.

In general we observe that conflict will occur if:
\begin{equation}
    \begin{split}
        & (r^{*}_{211}>r^{*}_{121} \land r^{*}_{122}>r^{*}_{212}) \\
        \lor & (r_{211}^{*}<r_{121}^{*} \land r_{122}^{*}<r_{212}^{*})
    \end{split}
    \label{eqn:conflict_definition}
\end{equation}
The definition of the AoC for the following decision-making models follows from Equation \ref{eqn:conflict_definition} (In order to save space we will let $A = r_{211}-r_{121}$ and $B = r_{122}-r_{212}$):

\begin{table*}[t]
    \centering
    \caption{Calculated AoC values for various interactive decision-making models based on the lane change reward matrix given in Figure \ref{intro_figure_table}, and general expressions for computing the AoC for each of the methods considered.}
    \label{AoC_specific_table}
    \begin{tabular}{lcc}
        \toprule
        Method                              & Lane Change AoC   & General AoC\\
        \midrule
        Baseline \citep{von1934}            & $1$               & 1 \\
        Pure Altruism \citep{andreoni1993}  & $1$               & $\min(\frac{A}{B},\frac{B}{A})$ \\
        SVO \citep{Schwarting2019}          & $0.5$             & $\frac{p_{1}p_{2} + (\frac{\pi}{2}-p_{1})(\frac{\pi}{2}-p_{2})}{(\frac{\pi}{2})^{2}}$ \\
        Altruism (Ours)                     & $0.5$             & $2(\frac{AB}{(A+B)^{2}})$ \\
        Aug-Altruism (Ours)                 & $0.38623$         & $\text{ln}(A+B)(\frac{A}{B} + \frac{B}{A}) - (\frac{A}{B}\text{ln}(A) + \frac{B}{A}\text{ln}(B)) - 1$  \\
        \bottomrule
    \end{tabular}
\end{table*}

\begin{figure*}[!ht]
    \centering
    \subfloat[\label{base_aoc}]{\includegraphics[width=.19\textwidth]{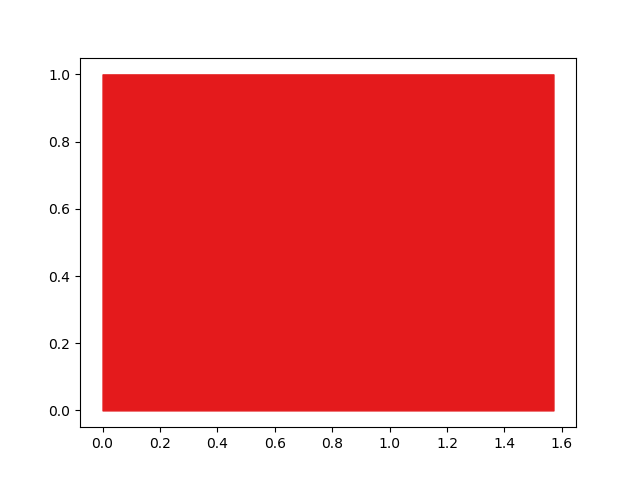}}
    \subfloat[\label{pure_alt_aoc}]{\includegraphics[width=.19\textwidth]{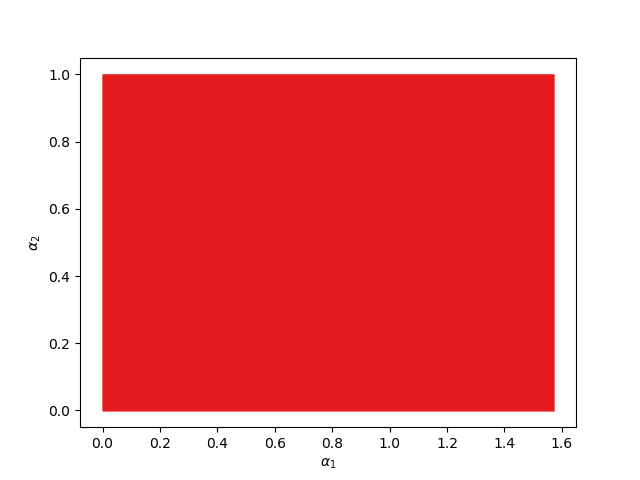}}
    \subfloat[\label{svo_aoc}]{\includegraphics[width=.19\textwidth]{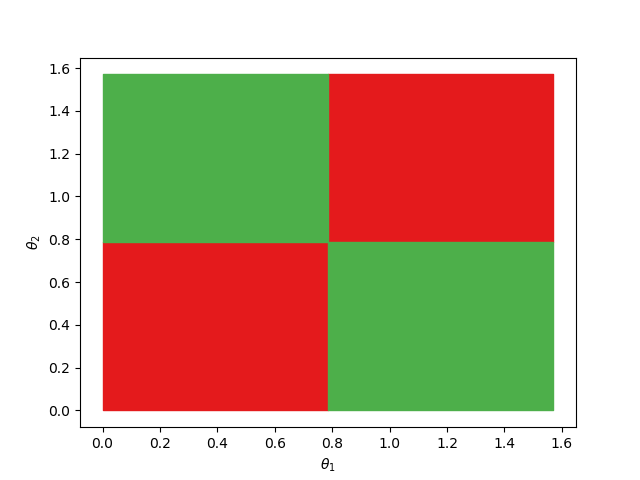}}
    \subfloat[\label{alt_aoc}]{\includegraphics[width=.19\textwidth]{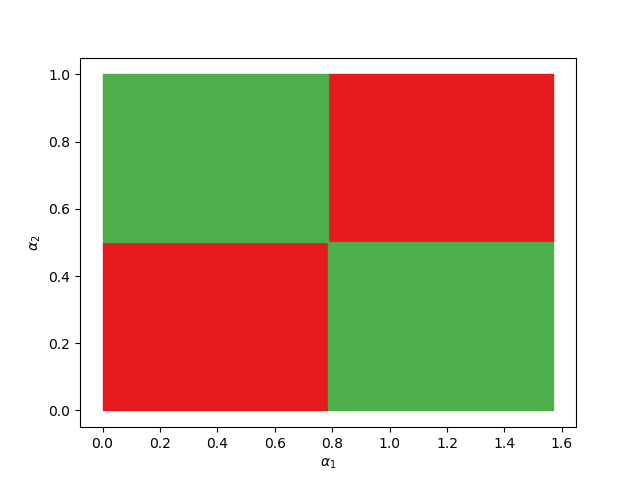}}
    \subfloat[\label{aug_alt}]{\includegraphics[width=.19\textwidth]{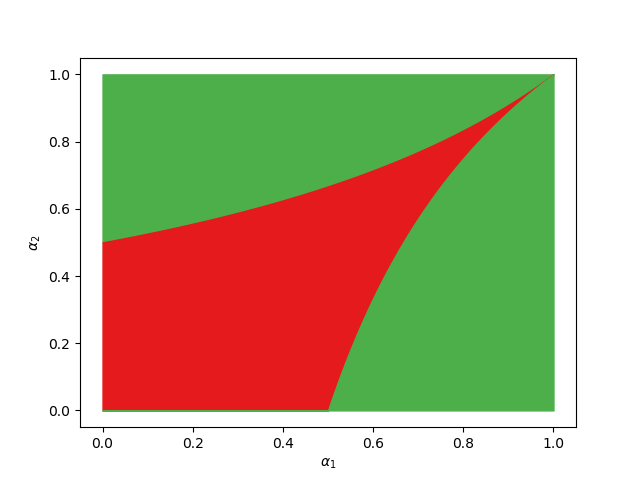}}
    \caption{Conflict regions for various reward models corresponding to the reward matrix in Figure \ref{intro_figure_table}. Red regions designate regions in the parameter space that result in conflict. Plots correspond to: (a) Baseline, (b) Pure Altruism, (c) SVO, (d) Altruism, (e) Augmented Altruism.}
    \label{constraint_bound_plot}
\end{figure*}

\begin{itemize}
    \item{Pure Altruism}: $(\alpha_{1}>\frac{A}{B} \land \alpha_{2}>\frac{B}{A}) \lor (\alpha_{1}<\frac{A}{B} \land \alpha_{2}<\frac{B}{A})$
    \item{Altruism}: $(\alpha_{1}>\frac{A}{B+A} \land \alpha_{2}>\frac{B}{B+A}) \lor (\alpha_{1}<\frac{A}{B+A} \land \alpha_{2}<\frac{B}{B+A})$
    \item{Augmented Altruism}: \\
            ($1-\frac{1-\alpha_{1}}{\alpha_{1}}\frac{A}{B}<\alpha_{2}<\frac{B}{B+(1-\alpha_{1}A)}$ $\land$  $0<\alpha_{1}<1$)
\end{itemize}

In the above it also holds that $0 \leq \alpha_{i} \leq 1$, except in the case of augmented altruism, where there is the extra constraint that $0<\alpha_{1}<1$. Each of the logical conjunctions ($\land$) specifies a bounded region of parameter space which will result in conflict, and the logical disjunctions ($\lor$) define pairs of non-overlapping regions (see Figure \ref{constraint_bound_plot} for a graphical depiction of these regions). Therefore we can define the AoC as the sum of the areas of these regions in parameter space. For Pure Altruism, and our proposed Altruism, these are straightforward computations. For the remaining derivations we refer the reader to Appendix \ref{deriving_area_of_conflict}. The general definitions for AoC of the standard Stackelberg Game, the Pure Altruism, SVO, Altruism, and Augmented Altruism variants are provided in Table \ref{AoC_specific_table} as well as evaluations corresponding to the reward matrix in Figure \ref{intro_figure_table}.

\begin{figure}[t]
    \begin{center}
    \subfloat[\label{a_vs_aoc_b_1}]{
        \includegraphics[width=.29\textwidth]{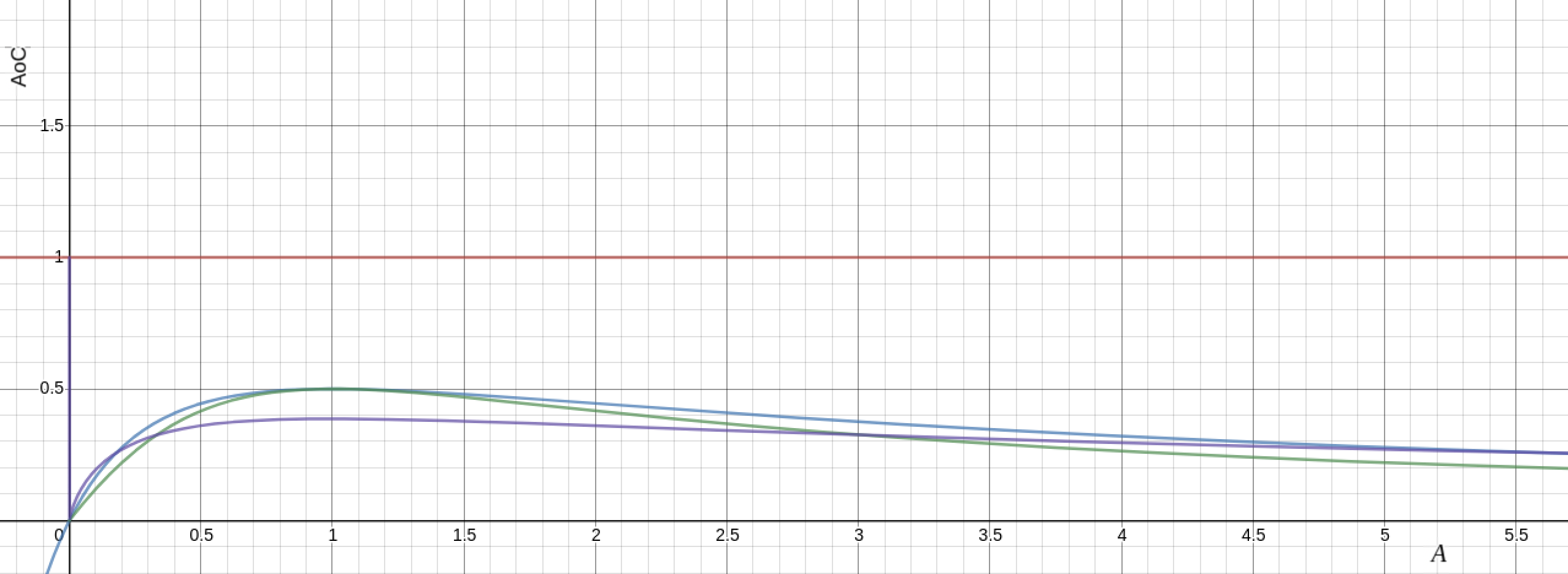}}
    \end{center}
    \begin{center}
    \subfloat[\label{a_vs_aoc_b_3_5}]{
        \includegraphics[width=.29\textwidth]{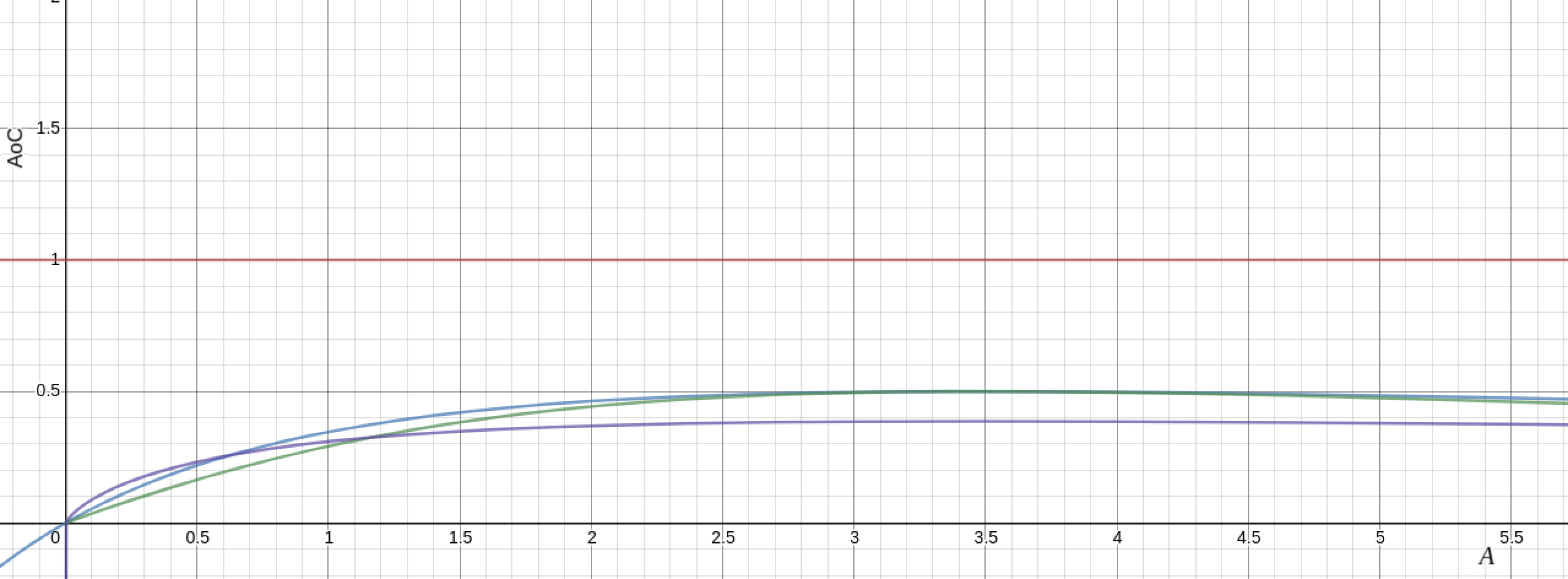}}
    \end{center}
    \caption{Plot of the AoC for varying values of $A$. The Blue line corresponds to Altruism, the Green SVO, and the Purple Augmented Altruism. (a) $B=1$, (b) $B=3.5$.  }
    \label{altruism_aoc_plots}
\end{figure}

We observe that the AoC for the Augmented Altruism significantly outperforms the other considered models. This means that, in repeated pairings of agents with altruism scores sampled uniformly from $[0,1]$, the incidence of conflict would be lowest when using this model. In general we empirically observe that, for reasonable magnitudes of $\frac{A}{B}$ Augmented Altruism consistently outperforms the other models. Figure \ref{a_vs_aoc_b_1} shows the AoC plotted against $A$, when $B=1$. We observe that for $0.33<A<3$ Augmented Altruism achieves minimal values. From Figure \ref{a_vs_aoc_b_3_5} we see that when $B=3.5$ this range is $1.6<A<10.4$. This demonstrates the effectiveness of the proposed model for minimising conflict.

\section{Human-AV Lane Change}
\label{human_av_lane_change}

The previous sections demonstrate that conflict can cause inefficient, and sometimes unsafe, execution of interactive manoeuvres. Our additional analysis shows that altruistic decision makers are able to reduce the incidence of conflict, but we have yet to show that human drivers behave altruistically---a modelling assumption that must be valid in order for our previous analysis to apply to real-world scenarios.

In order to verify this assumption, we gathered data from human participants interacting with an autonomously controlled vehicle. In total $11$ participants were recruited for the experiment, collectively contributing more than 1,400 trajectory demonstrations. We investigate the consistency of the observed data with two models of decision making: a conventional Stackelberg game that does not take altruism into account, and the decisions that would be made by an agent exhibiting \emph{egoistic} altruism. We hypothesise that data will support the egoistic altruism model.

\subsection{Experimental Setup}
Participants control a simulated vehicle in a highway lane change scenario, similar to the scene presented in Figure \ref{intro_figure_image}; The experiment setting has no obstacle vehicle, as in the figure. Instead participants are directed to complete the lane change as early as possible, and the environment has fixed length of $100$ metres, simulating the urgency inherent to the presented scenario. Each scenario starts with the cars next to each other driving at $5m/s$ in adjacent lanes. A scenario ends when both cars end up in the same lane travelling in the lane direction. The speed of either vehicle cannot exceed $5.5m/s$.

The participants control their vehicle using keyboard arrow-based controls. Participant input is in the form of action pairs $u=(a,\omega) \in [-3m/s^2,3m/s^2] \times [-10^{\circ}/s^2,10^{\circ}/s^2]$. Before beginning the experiments participants have access to a sandbox environment enabling them to familiarise themselves with the controls. This environment is a highway setting, identical to the experiment setting. The participant vehicle starts in the left lane travelling at $5m/s$. An obstacle car is stationary halfway along the road in the other lane. This allows the participant to practise merging manoeuvres in proximity with another vehicle. The simulation repeats until the participant is ready to continue on to the experiments.

In the first experiment, participants control the lane-changing vehicle, and are directed to change lanes as quickly as possible. In the second scenario the participants control the lane-keeping vehicle, facilitating a lane change by an autonomously controlled vehicle. The aim of the AV is to successfully complete the lane change as quickly as possible.

In all experiments the AV is controlled by a rule-based controller. When the AV is performing lane-keeping, it initially follows a constant velocity model. If the participant car attempts to enter the lane ahead of the AV, it changes models to an Intelligent Driver Model (IDM) controller (\citep{Treiber2000}). The model will either produce ``aggressive'' or ``passive'' behaviour; the ``aggressive'' model has the AV accelerate to block the participant's lane change, and the ``passive'' model has the AV slow down and make ample space to complete the manoeuvre. The parameters of both models are empirically chosen to demonstrate these behaviours. If the participant car does not attempt to cut in ahead of the other vehicle, then the AV model remains a constant velocity model. In the lane changing case the AV's trajectory is defined by polynomial curves. Initially the car follows a ``standard'' curve that performs a gradual lane change. If the participant does not yield, the AV will either aggressively adopt a faster lane change trajectory, or else it will give up on the lane change and merge behind the participant instead. The participant is not informed of the ``type'' of the AV during the experiment, and the ``type'' is randomly sampled at the start of each scenario such that each participant experienced each AV type an equal number of times. Each scenario took about $20$ seconds to complete.

It is clear from the definition that this formulation matches the game defined in Figure \ref{intro_figure_table}; the lane-changing vehicle wants to perform the lane change as quickly as possible, requiring cutting ahead of the lane-keeping vehicle. So the preferred equilibrium is $(LCA,Y)$. The lane-keeping vehicle can accelerate to get out of the way of the lane-changer, which will complete the manoeuvre quicker, preferring the $(LCB,C)$ equilibrium. No other outcomes are desirable.

\subsection{Sub-objectives and Altruism}
\begin{figure}[t]
    \centering
    \subfloat[\label{fig:h_exp1a_results}]{%
        \includegraphics[width=.5\columnwidth]{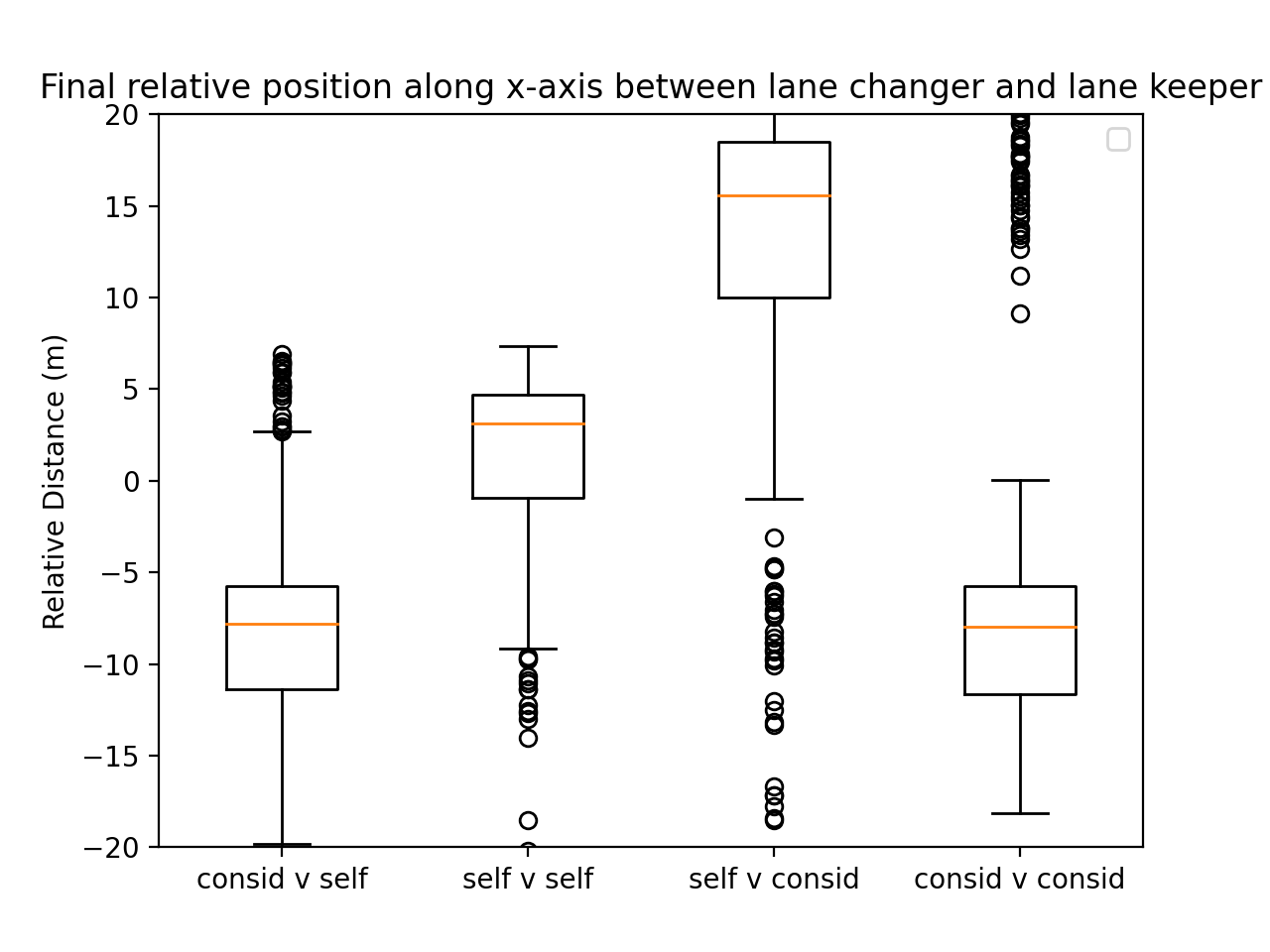}}
    \subfloat[\label{fig:h_exp1b_results}]{%
        \includegraphics[width=.5\columnwidth]{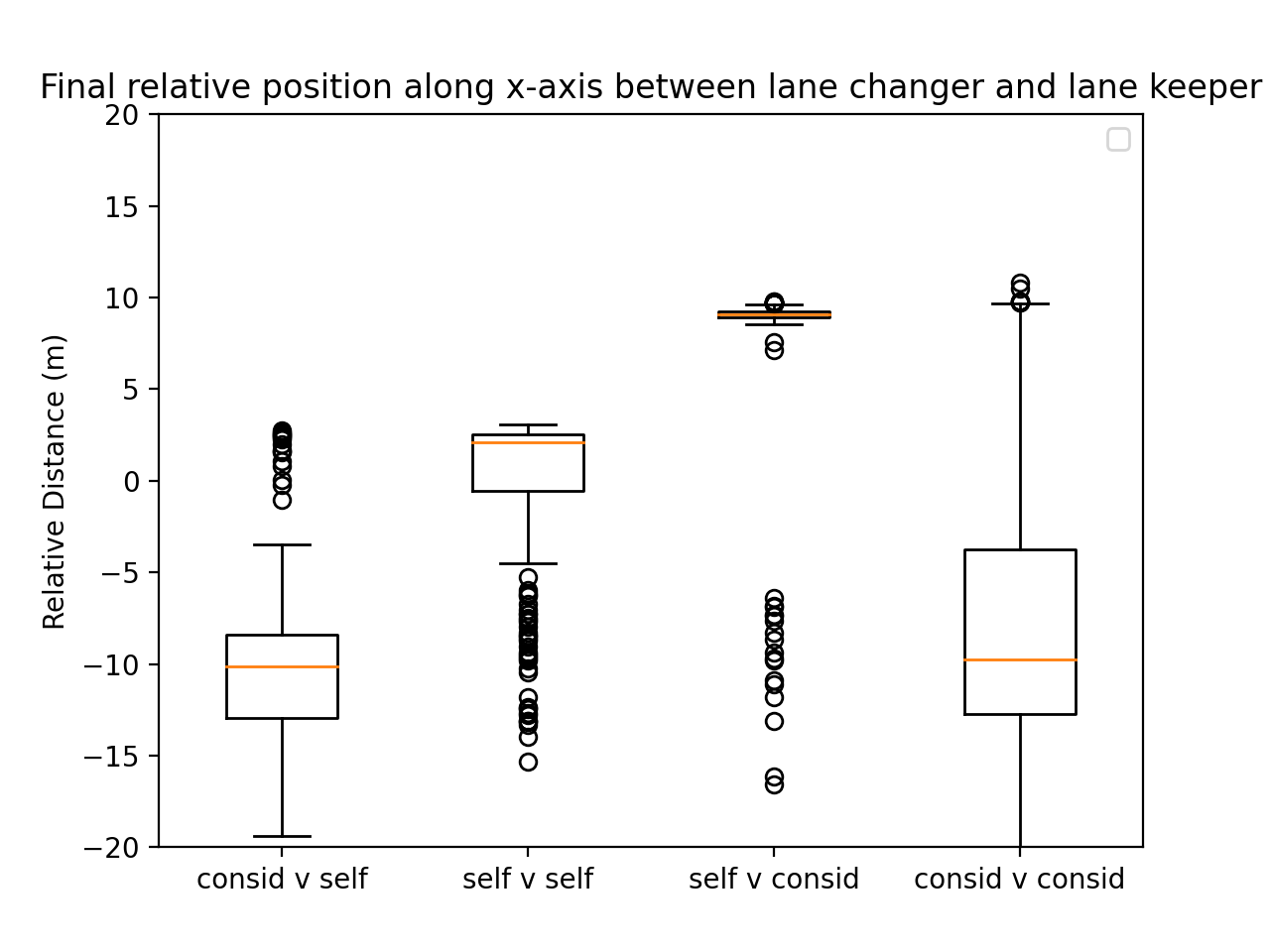}}
    \caption{Final relative displacement of the participant car with respect to the autonomous car. A positive value indicates the participant ended up ahead of the AV, and a negative behind. Cars within $5$ metres of each other have crashed; a) Results for the participant lane-change experiment; b) Results for the participant lane-keeping experiment.}
\label{sub_objective_results}
\end{figure}

Before each scenario the participants are given a ``sub-objective'' indicating how they are to complete their objective (lane-change or lane-keep). The possible sub-objectives are to drive:
\begin{itemize}
    \item ``as if in a rush'',
    \item ``considerately''.
\end{itemize}

Participants are told that the AV also has its own sub-objective that would affect its behaviour, but they are not told what the sub-objective is. By withholding this information from the participant we remove their ability to reason about the altruism that could be exhibited by the AV, thus precluding the user from performing non-egoistic altruism. The purpose of the sub-objective is to induce selfish (low-altruism) and altruistic (high altruism) behaviour demonstrations. The ``type'' of the AV controller, as defined previously, corresponds to the assigned sub objective. By having the AV's behaviour depend on this sub-objective, we can observe instances of conflict in human interactions with an autonomously controlled vehicle. Sub-objective pairs were sampled randomly such that each combination was observed $15$ times per experiment.

Figure \ref{sub_objective_results} gives the final relative displacement of the participant controlled vehicle with respect to the AV at the end of each scenario. We can see that when the participant was given the ``considerate'' objective (``consid v self'' and ``consid v consid'' columns), they consistently allow the AV to complete the manoeuvre ahead of them. Based on Figure \ref{intro_figure_table}, this is what we would expect to be the result of a driver with a high altruism coefficient. We observe that when the participant was directed to behave selfishly, and the AV was being considerate (``self v consid''), the participant consistently ended up ahead of the AV. This is what we would expect from a low-altruism participant and a high altruism AV.

We observe that conflict is also evident from these results. In the case when both cars behave selfishly ("self v self"), the cars crash into each other, as both prioritise completing their objective over the other vehicle's. This aligns with what we would expect from two cars with low altruism coefficients.\footnote{We note that we do not observe conflict in the case where both cars are passive since, for technical reasons, the autonomous car's behaviour only changes if the participant is ahead of them. Therefore the mutual yielding equilibrium is never realised.} Based on these results we conclude that the directions provided to the human participants reliably induced demonstrations of low-altruism and high-altruism behaviour in human participants. Collectively, these results support the hypothesis that egoistic altruism-based models (e.g., altruism, pure altruism, SVO) can be used to approximate human decision-making behaviour more accurately than conventional Stackelberg games.

\section{Augmented Altruism in Human Decision-making}


We have shown that Augmented Altruism achieves a lower AoC score than alternative approaches. While the previous experiment demonstrates that humans can behave according to egoistic interactive models when driving, it remains to be shown that \emph{non-egoistic} interactive models can capture the decision-making done by humans decision-makers more reliably than egoistic models. In \citep{Schwarting2019} it was shown that the SVO-based model sufficed to capture human decision-making in a highway setting. In this work we have shown that our standard Altruism model is equivalent to the SVO-based model, both of which are egoistic interactive models.

Our hypothesis when defining the Augmented Altruism coefficient was that human decision-making depends not only on the planning agent's altruism coefficient, but also their estimate of the coefficient of the other agent. In this section we will validate this hypothesis using results from an experiment run with human participants driving with an autonomously controlled vehicle. Specifically, we show that non-egoistic behaviour is more consistent with the observed data than egoistic behaviour.

\subsection{Experimental Setup}
This experiment immediately followed the completion of the experiments in Section \ref{human_av_lane_change}. At the start of the experiment each participant is informed that the experiment they had just completed was to do with learning aggressive and passive behaviour. They are told that, using the demonstrations they had provided, a model of their behaviour had been learnt, and that this model could be made more or less aggressive as required.

The experiment consists of four scenarios, identical to the previous setup, with the participant performing a lane change manoeuvre. Before each scenario the participant is given one of the following notifications:
\begin{itemize}
    \item The AV would behave more aggressively than their demonstrated behaviour,
    \item The AV would behave less aggressively than their demonstrated behaviour,
    \item The AV would behave more passively than their demonstrated behaviour,
    \item The AV would behave less passively than their demonstrated behaviour,
\end{itemize}

The order in which these were presented to the participant was randomised, and each participant was shown each notification once. The resulting behaviour of the participant vehicle was recorded.

In reality the AV's model was the same for all the scenarios, following the passive rule-based IDM controller specified earlier. The only parameter varied in the experiment was the participant's belief about the altruistic tendency of the AV.

\subsection{Results}
\begin{figure}
    \centering
    \includegraphics[height=.2\textheight]{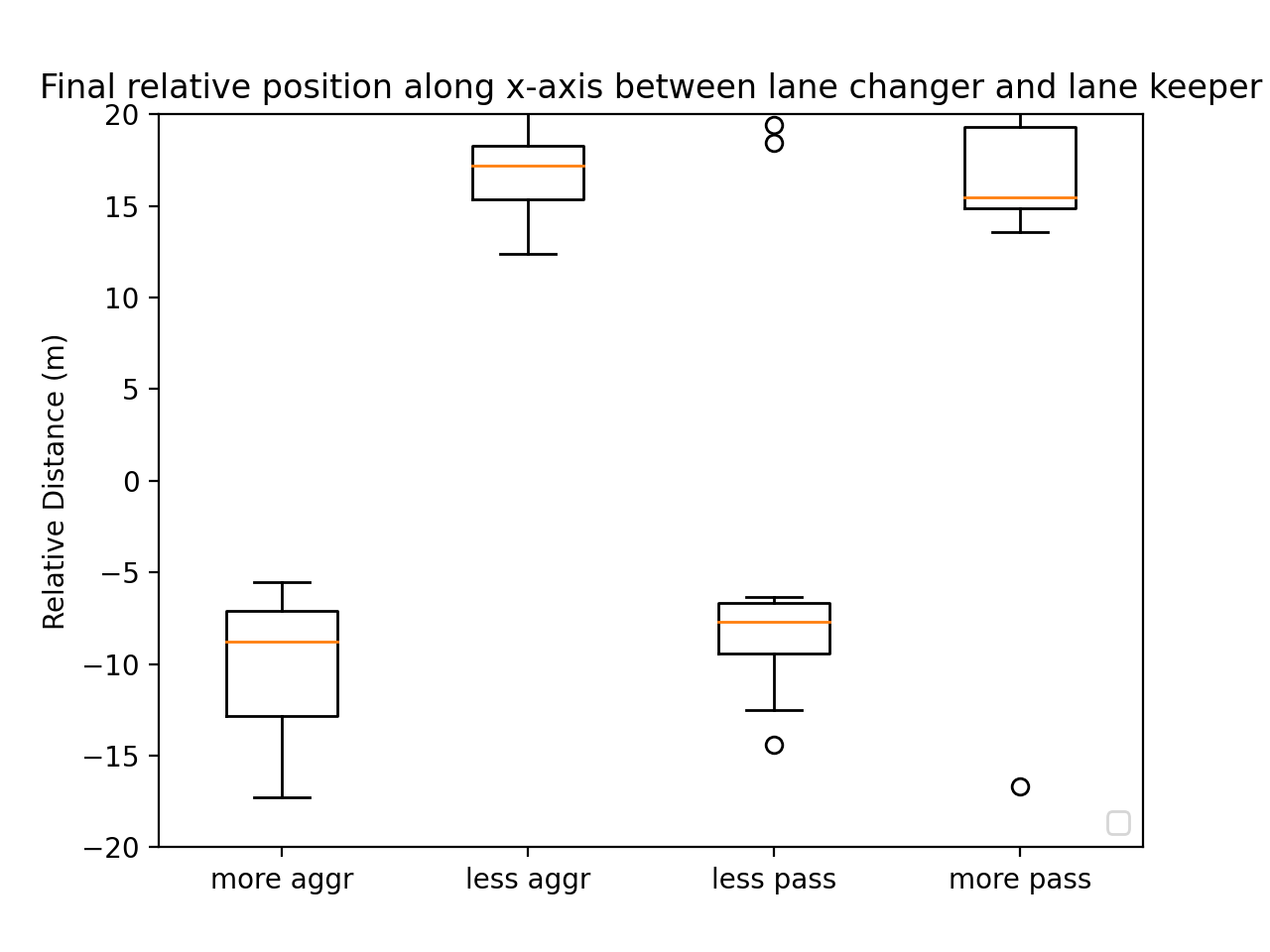}
    \caption{Final relative displacement of the participant car with respect to the AV. A positive value indicates the participant ended up ahead of the AV, and a negative behind. The columns are labelled as follows; more/less aggr: scenarios when participants were told the AV was more/less aggressive than them; more/less pass: scenarios when participants were told the AV was more/less passive than them.}
    \label{fig:h_exp2_results}
\end{figure}

Figure \ref{fig:h_exp2_results} gives the results from this experiment. If the participants were performing decision-making purely egoistically, per the standard altruism model, then we would expect to see no change in behaviour resulting from the different instructions, as these instructions only pertained to the value of the AV's altruism coefficient. However, as we see, distinct behaviours are apparent; when informed they were less aggressive (or more passive) than the other car, participants consistently elected to merge behind the AV. When they were informed of the opposite they chose to merge ahead of the vehicle. Crucially, this behaviour does not match the decision-making that would be predicted by either SVO or standard Altruism. Because information about the AV's behaviour changes the trajectory chosen by the participant, we conclude their model must depend on both the agent's altruism coefficient, as in the Augmented Altruism model, thus supporting our hypothesis.

\section{Conclusion}
In this work we identified conflict, a vulnerability in popular decision methods for autonomous driving. We experimentally demonstrated the significant negative consequences conflict can have in common interactive driving scenarios. We proposed a metric, Area of Conflict, that measures the potential incidence of conflict for a decision-making method. We derived a novel method for adjusting the values used in decision-making in such a way to reduce the method's vulnerability, and provided theoretical guarantees that the methods reduce the incidence of conflict for reasonable reward values.
\section*{Acknowledgements}
Jack Geary is supported by a postgraduate studentship sponsored by FiveAI Ltd.

\bibliographystyle{plainnat}

\bibliography{references}

\newpage
\appendix
\section{Appendices}
\subsection{Deriving Augmented Altruism}
\label{deriving_augmented_altruism}
When attempting to identify equilibria in Game Theoretic problem formulations it is not uncommon to used iterative best response methods to compute the Nash Equilibrium \citep{vorobeychik2008}. In practise this involves each agent choosing an optimal action based on the optimal actions for the other agents in the previous iteration. This allows for the fact that an agent's choice of action can be affected by the choices made by other agents. If this process is repeated indefinitely, and it converges to a solution, then the solution achieved is a Nash Equilibrium \citep{bacsar1998}. 

The altruism definition presented in Equation \ref{eqn:altruism_definition} neglects from consideration that $r_{-i}^{*}$ is the reward an altruistic agent $-i$ would receive from the interaction, not $r_{-i}$, and that awareness of this value could affect agent $i$'s preferences. But, by the same assumption, the value for $r_{-i}^{*}$ depends on the value of $r_{i}^{*}$.Therefore, in order to adequately account for the altruistic inclinations of both agents, we use iterative methods over the system of equations in Equation \ref{eqn:altruism_definition}. We observe that repeated iteration over the system of equations allows each agent to account for the other's altruistic coefficient; after each agent computes their altruistic reward once, they can repeat the process using the rewards computed in the previous iteration. This gives us the following system of equations:

\begin{equation}
    \begin{split}
        r_{1}^{k}  &= (1-\alpha_{1})r_{1} + \alpha_{1}r^{k-1}_{2} \\
        r_{2}^{k}  &= (1-\alpha_{2})r_{2} + \alpha_{2}r^{k-1}_{1},
    \end{split}
\end{equation}
where $k\geq 0$ gives the iteration index. Agent $i$ does not iterate over reward $r_{i}$ as the amount of reward they would get from achieving their own objective, $(1-\alpha_{i})r_{i}$, is known and does not need to be optimised. Since the altruism coefficients are bounded, $0 \leq \alpha_{i} \leq 1$, we know this system will converge (provided $\alpha_{1}$ and $\alpha_{2}$ are not both exactly $1$, as this renders the computation unsolvable). We can find the steady state for this system by solving:

\begin{equation}
    \begin{split}
        r_{1}^{\infty} &= (1-\alpha_{1})r_{1} + \alpha_{1}r^{\infty}_{2} \\
        r_{2}^{\infty}  &= (1-\alpha_{2})r_{2} + \alpha_{2}r^{\infty}_{1} \\
    \end{split}
\end{equation}

This solution gives the definition of the altruistic reward presented in Equation \ref{eqn:augmented_altruism_definition}.
\begin{equation}
    r_{i}^{*} = \dfrac{(1-\alpha_{i})r_{i}+\alpha_{i}(1-\alpha_{-i})r_{-i}}{1-\alpha_{i}\alpha_{-i}} \quad i \in \{1,2\}
\end{equation}

\section{Deriving Area of Conflict}
\label{deriving_area_of_conflict}
We recall that conflict will occur for Augmented Altruism if:
\begin{equation}
    (1-\frac{1-\alpha_{1}}{\alpha_{1}}\frac{A}{B}<\alpha_{2}<\frac{B}{B+(1-\alpha_{1}A)} \land  0<\alpha_{1}<1)
\end{equation}
By solving these inequalities we get the following definition for AoC (With $A = r_{211}-r_{121}$ and $B = r_{122}-r_{212}$):

\begin{equation}
    \begin{split}
        AoC & = \int_{0}^{1}\frac{B}{B+(1-\alpha)A}d\alpha - \int_{\frac{A}{A+B}}^{1} 1-\frac{1-\alpha}{\alpha}\frac{A}{B} d\alpha \\
        &= ln(A+B)(\frac{A}{B}+ \frac{B}{A})- (\frac{A}{B}ln(A) + \frac{B}{A}ln(B)) - 1
    \end{split}
\end{equation}\ 

For comparison we can also perform the same evaluation for SVO \citep{Schwarting2019}.

\begin{equation}
    r_{i}^{*} = \cos(\theta_{i})r_{i} + \sin(\theta_{i})r_{-i} \quad 0\leq \theta_{i} \leq 2\pi 
\end{equation}\

By the same procedure as before we observe that conflict occurs with SVO when:
\begin{equation}
    \begin{split}
    &(\theta_{1}<\tan^{-1}(\frac{A}{B}) \land \theta_{2}<\tan^{-1}(\frac{B}{A})) \\ 
    \lor & (\theta_{1}>\tan^{-1}(\frac{A}{B}) \land \theta_{2}>\tan^{-1}(\frac{B}{A}))
    \end{split}
\end{equation}\

Even though the SVO mechanism allows for masochistic and sadistic behaviours (corresponding to angles resulting in coefficients with negative magnitudes), to facilitate comparison we constrain the SVO coefficients to be between 0 and 1. This  implies $0<\theta_{i}<\frac{\pi}{2}$. We can therefore compute the AoC for SVO as:

\begin{equation}
    \begin{split}
        p_{1} &= max(0,min(\frac{\pi}{2},\tan^{-1}(\frac{A}{B}))) \\ 
        p_{2} &= max(0,min(\frac{\pi}{2},\tan^{-1}(\frac{B}{A}))) \\
        AoC &= \frac{p_{1}p_{2} + (\frac{\pi}{2}-p_{1})(\frac{\pi}{2}-p_{2})}{(\frac{\pi}{2})^{2}}
    \end{split}
\end{equation}

\end{document}